\RequirePackage{fix-cm}
\documentclass[twocolumn]{svjour3}          
\smartqed  
\usepackage{graphicx}
\usepackage[version=4]{mhchem}
\usepackage{textcomp}
\usepackage{authblk}
\usepackage{lineno}
\begin{document}

\title{Investigating the effect of different quantum dots on the absorption spectrum and characteristics of quantum dot sensitized solar cells
}


\author{Hossein Vahid Dastjerdi         \and Hamidreza Fallah \and Morteza Hajimahmoodzadeh
}


\institute{F. Author \at
              Hossein Vahid Dastjerdi \\
              \and
		   S. Author \at
		   	  Hamidreza Fallah \\             
              \email{hfallah@sci.ui.ac.ir}          
          \and
           t. Author \at
              Morteza Hajimahmoodzadeh
}


\maketitle

\begin{abstract}
Quantum dot sensitized solar cells are among the new generations of solar cells that have attracted much attention. Theoretical and simulation studies have predicted high efficiency for these cells so that in the future, these cells could be an excellent alternative to silicon solar cells. Other advantages of these cells are their ease of fabrication and cheaper manufacturing methods than existing cells. This paper's main idea is to simulate the effect of different quantum dots on the optical and electrical characteristics of these cells and, in particular, the efficiency. We then simulated the effect of simultaneous sensitizing by different quantum dots, and we observed that the cell's light absorption and the efficiency in simultaneous sensitizing, increased. Then we experimentally studied one of the cells that give the best simulation result (\ce{PbS}/\ce{CdS} co-sensitized). We deposited the quantum dots on transparent \ce{TiO2}, and we obtained the light absorption, efficiency, and other characteristics of cells. Further, we investigated the effect of cobalt sulfide as the counter electrode in \ce{PbS}/\ce{CdS}, instead of platinum and gold, and we found that the efficiency has increased.

\end{abstract}

\section{Introduction}
\label{intro}
can be transformed into other useful forms of energy, such as heat, electricity, and chemicals, to supply daily human energy. Reports have shown that using solar cells by 10\% efficiency in just 0.1\% of the earth's surface can provide current electricity demand \cite{gratzel2001photoelectrochemical}. The first-generation solar cells were based on silicon, and then the thin-layer solar cells were made as second-generation. The production of solar cells with high efficiency, low cost, and mass-producing capability is a priority of research in solar cells. Researchers have developed a plan for the third generation of solar cells by combining the knowledge of first and second generations of solar cells. This generation can achieve Shockley's extraordinary theoretical efficiency while reducing cell manufacturing costs \cite{beard2013third}.

Quantum dot solar cells are a bunch of solar cells based on nanomaterials, and quantum dots(QDs) and QDs can be used in a variety of solar cells to reduce costs and increase efficiency. The optical behavior of quantum dots is such that they emit visible light of varying wavelengths when exposed to ultraviolet light \cite{els1}. The point is that the wavelength of light emitted from quantum dots depends on the quantum dots' size. In small quantum dots, the band-gap is larger. Therefore, by applying the UV beam to small quantum dots, the electrons traveling to the higher energy band emit a larger energy band-gap when losing excess energy and return to steady-state. The beam of visible light they emit has more energy and is bluish. Quantum dots have attracted researchers' attention to making solar cells because of their adjustable band-gap to their desired size. So their absorption spectra can be adjusted to the spectral distribution of sunlight \cite{els2}. Their high extinction coefficient \cite{els3}, rapid separation of charges due to the significant intrinsic dipole moments of them \cite{els4}, and their ability to produce multiple excitons by absorbing a single photon cause the incident photon to current conversion efficiency to be more than 100\%~\cite{els5}.

If the photon energy exceeds the threshold required for impact ionization, this extra energy will excite another electron and produce another exciton \cite{wolf1998solar}.  Impact Ionization is a process in which an electron or hole with sufficient kinetic energy can excite one electron from the valence band to the conduction band. The result of this process is the production of another electron-hole pair.

Quantum dot sensitized solar cells (QDSSCs) are a replacement of dye-sensitized solar cell (DSSCs) \cite{mishra2009metal}. The difference between QDSSCs and DSSCs is sensitizers which are responsible for absorbing light. In DSSCs, the sensitizers are organic dye molecules or metal-organic compounds, and in QDSSCs, they are quantum dots. Briefly, the operation principles of these cells are as follows. Li research group in 2013 showed that using \ce{PbS} and \ce{CdS} quantum dots deposited on \ce{TiO2} nanorod arrays can achieve to 1.3\% efficiency \cite{li2013efficient}. In this work, we improved the efficiency of cells using transparent and reflector \ce{TiO2} to 1.88\% . 
 
After the light is emitted into the photovoltaic cell, the quantum dots on the surface \ce{ TiO2} layer absorb light, which excites the quantum dots' electrons from the valance band to the conduction band. This process leaves a positive charge hole in the vacancy of the electron. The generated electron-hole pair (exciton) must be separated at the quantum dots' boundary with \ce{ TiO2}. The excited electron is transferred to the \ce{TiO2} electrode and moved to the counter electrode through an external circuit. Also, through the electrolyte, which plays the role of oxidation /reduction (redox) pair, the electron transfers to the Ground state of the quantum dots and becomes oxidized. Then the intermediate undergoing the oxidation process goes through the diffusion process to the counter electrode and is then subjected to the reduction process again.

Selecting the counter electrode is one of the key points in achieving high efficiency. Because the counter electrode plays the role of oxidation, it is necessary to choose the counter electrode from materials with low resistance and high work function. Mainly in quantum dot sensitized solar cells, the $\ce{S}^{2-} / \ce{S_{x}^2-}$ is redox pair, and the platinum, gold, and cobalt sulfide are used as the counter electrode. We used three of them in \ce{PbS}/\ce{CdS} co-sensitized solar cells, and we investigated their effect on the efficiency of this particular cell in the experimental section.  

In the simulation section of this paper, after investigation the effect of different quantum dots on the optical and electrical properties of sensitized solar cells with one or two types of QDs, we optimized the size of QDs for the cell with the highest predicted efficiency. As we describe in the Experimental section, we fabricated this solar cell in the laboratory and investigated the effect of different electrodes on its efficiency.

\section{Simulation method}
\label{sec:1}
If the wavelength of the incident light is less than or equal to the size of the structure, we will no longer be able to use the ray optics analysis methods for obtaining accurate solutions. Therefore, we will need a new optical approach to provide the correct and expected response. For optical simulation, we used the method Finite-difference time-domain (FDTD) because this method can solve Maxwell’s equations directly in the time domain, thereby providing accurate broadband solutions to electromagnetic wave propagation and scattering problems \cite{ccapouion}. FDTD is a vector-based approach that provides users with information about frequency and time domains and provides a different perspective on issues and applications in electromagnetism and photonics. The FDTD method has been one of the successful methods that do not use matrix inversion, and due to its purely computational nature, the FDTD is free from linear algebra problems that limit other frequency domain analysis methods to one million unknowns \cite{chen2004electrical}. Hence, models with one billion unknowns are implemented for FDTD, and in theory, there is no high limit on the number of unknowns \cite{chen2004electrical}. The sources of error in the FDTD method are well known and can be reduced for more accurate answers. Therefore, only a single run of the simulation can achieve a system frequency response over a wide range. The FDTD method is a systematic one, and the analysis of a new structure is reduced to a networking problem by this method, and no need to rewrite complex integral equations \cite{tafloval}.

As the name implies, the time-domain finite difference method is done in the time domain. So when a simulation is run, the Maxwell equations are, in fact, solved in the time domain to obtain E (t) and H (t). In FDTD Method we solve Maxwell's equations by dividing the electric and magnetic fields, which are initially continuous functions of time, and obtain the fields in time $n\Delta t$ that we briefly represent with n using step-time equations (see Eqs (\ref{5eq13}) and (\ref{5eq14}) \cite{xu2014photonics}).

\begin{align}
\vec{E}^{(n+1)}=\vec{E}^{(n)}+\frac{\Delta t}{\varepsilon}\vec{\nabla}\times \vec{H}^{(n+\frac{1}{2})}\label{5eq13},\\
\vec{H}^{(n+\frac{3}{2})}=\vec{H}^{(n+\frac{1}{2})}-\frac{\Delta t}{\mu}\vec{\nabla}\times \vec{E}^{(n+1)}\label{5eq14},
\end{align}
\noindent Where $\varepsilon$ is the electrical permittivity, and $\mu$ is magnetic permeability. Using Eqs (\ref{5eq13}) and (\ref{5eq14}) and having the primary electric field, we can calculate the final value of the electric and magnetic fields. But most of the time we are looking for an electric field in terms of frequency. A single-phase or continuous-wave field in a steady-state should be obtained from the electric field in terms of the Fourier transform time during the simulation as Eq.(\ref{eq100}) \cite{xu2014photonics}. Consequently, by taking the Fourier transform of the electric field in the time domain during the simulation, the electric field will be obtained at any particular frequency or equivalent in any specific wavelength.

\begin{equation} 
\vec{E}(\omega)=\int_0^T e^{i\omega t} \vec{E}(t) dt. \label{eq100}
\end{equation}

Eq.(\ref{eq2}) can also be used to calculate the power absorbed by the cell per unit volume per frequency \cite{gross2016theoretical}. To calculate it, we only need to know the electric field intensity and the imaginary part of the electrical permittivity of the material.

\begin{equation}
P_{abs}(r,\omega)=-0.5\times\omega |{E}|^{2}Im(\varepsilon).\label{eq2}
\end{equation}

The number of photons absorbed per unit volume per frequency can be calculated by dividing the absorption power in that frequency by the energy of each photon in the same frequency by Eq. (\ref{eq12s}) \cite{xu2014photonics}.
\begin{equation}
g=\dfrac{P_{abs}}{ \hbar \omega}=\dfrac{-0.5\times |{E}|^{2}Im(\varepsilon)}{\hbar}. \label{eq12s}
\end{equation}
If we assume each absorbed photon produces an electron-hole pair, the rate of electron-hole production $ (G) $  can be obtained by integrating Eq. (\ref{eq12s}) over the entire simulated spectral region \cite{xu2014photonics}.
\begin{align}
G=\int_{solar\hspace{1mm}spectrum}{g d\omega} \\
=\int_{solar\hspace{1mm}spectrum}\dfrac{-0.5\times |{E}|^{2}Im(\varepsilon)} {\hbar} d\omega.
\end{align}
By solving the equations of drift and diffusion for carriers, given separately for electrons and holes by  Eqs.(\ref{eq3}) and (\ref{eq4}), and by considering the continuity equations, the production rate and the recombination rate for each carrier, we can obtain a voltage-current diagram of solar cells \cite{xu2014photonics}. 
\begin{align}
J_{n}=q\mu_{n}nE+qD_{n}\nabla{n}\label{eq3},\\
J_{p}=q\mu_{p}pE-qD_{p}\nabla{p}\label{eq4},
\end{align}
\noindent where $q$ represents the electron's charge, $\mu_{n}$ and $\mu_{n}$ are the mobility of electrons and holes, $n$, and $p$ are the densities of electrons and holes, respectively. $ E$ is the electric field, $D_n$ and $D_p$ are the diffusion coefficients of electrons and holes, respectively.
Efficiency can then be calculated using Eq.(\ref{eq10}) \cite{luque2011handbook}:

\begin{equation}
\eta = \dfrac{J_{SC}\times V_{OC}\times FF}{P_{in}}, \label{eq10} 
\end{equation} 
\noindent where $P_{in}$ is the power intensity of the incident light (100 $mW/cm^{2}$) and Fill Factor $F\hspace{-0.5 mm}F$ is defined by Eq. (\ref{eq101}) \cite{luque2011handbook}.
\begin{equation} 
FF = \dfrac{P_{max}}{J_{SC}V_{OC}}, \label{eq101}
\end{equation}
\noindent where $P_{max}$ is the maximum power of a photovoltaic cell, $J_{SC}$ is short circuit current density and $V_{OC}$ is open circuit voltage.
\subsection{The purpose of simulating solar cells}
With the goal of reducing production costs and increasing the efficiency of solar cells, research on solar cells is increasingly focused on new solar cell design concepts, including nanostructured solar cells. Solar cell simulation is necessary to predict the behavior of these devices. As the complexity of the design of photovoltaic cells increases, it becomes difficult to obtain analytical solutions to describe their performance. In a real cell, non-ideal processes such as surface and volume recombination of carriers reduce the efficiency of solar cells. A combination of optical and electrical simulations that considers these non-ideal processes is essential to simulate solar cells more accurately.
\subsection{Simulation of the effect of quantum dot type on absorption and electrical characteristics of QDSSCs}
\begin{figure}
\centering
\includegraphics[scale=.42]{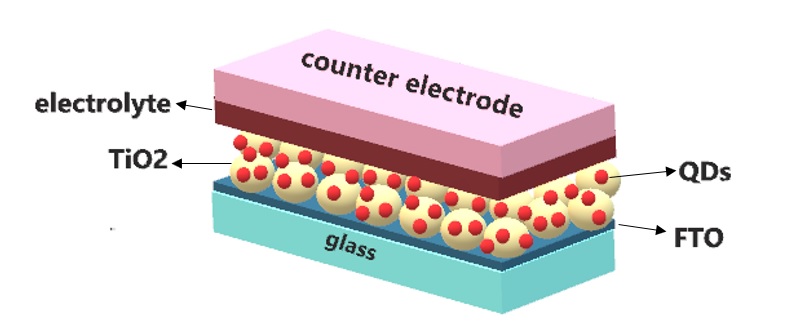}
 \caption{Schematic structure of a quantum dot sensitized solar cell.\label{p12}}
\end{figure}

The usual structure of these cells is \ce{FTO} / \ce{TiO2} / \ce{ QDs} / \ce{electrolyte} / \ce{counter electrode} as shown in the Fig.~\ref{p12}. The active area of these cells includes \ce{TiO2} and quantum dots. Quantum dots are the major absorbers of sunlight and excitons are produced after absorbing light using QDs. Photocurrent is formed by the transfer of electrons to the \ce{TiO2} anode \cite{lee2016electron}. Based on the previously described relationships, we simulated the solar cells with different sensitizers. 

\begin{figure}
\centering
\includegraphics[scale=.5]{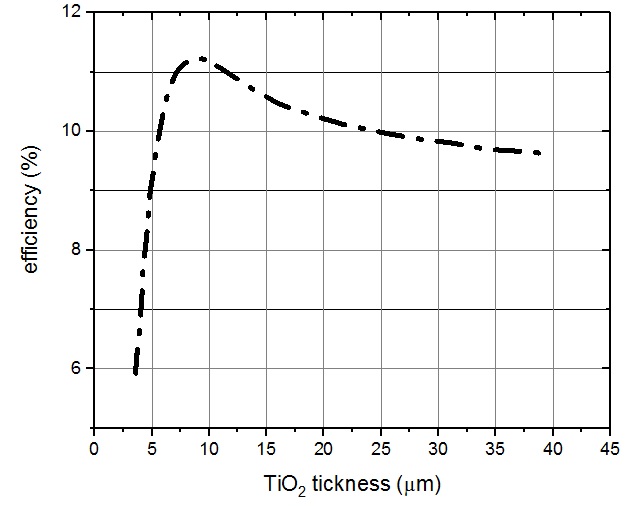}
 \caption{Cell efficiency changes based on \ce{TiO2} layer thickness.\label{p13}}
\end{figure}

For quantum dot sensitized and dye-sensitized solar cells, the best \ce{TiO2} layer thickness is $10 ~\mu m$ \cite{thicktio}. We optimized the \ce{TiO2} layer thickness for all QDSSCs, which we discuss in the simulation section, to get the most efficiency. The result is shown in Fig.~\ref{p13}, which confirms the access to the highest efficiency at approximately $10 ~\mu m$.

Although the open circuit voltage decreases with increase in thickness of \ce{TiO2} layer due to increase in electron diffusion length to the electrode, the short-circuit photocurrent density ($J_{SC}$) increases with film thickness due to enlargement of surface area \cite{domtau2017effects}. When the thickness of the \ce{TiO2} layer increases over an optimum quantity, the carrier recombination rate increases and the short circuit current drops, Which has a negative effect on cell efficiency. In the simulation, we consider the layer thickness $10 ~\mu m$ and discuss the effect of different quantum dots as sensitizer on the cell characteristics.

\begin{figure}
\centering
\includegraphics[scale=.5]{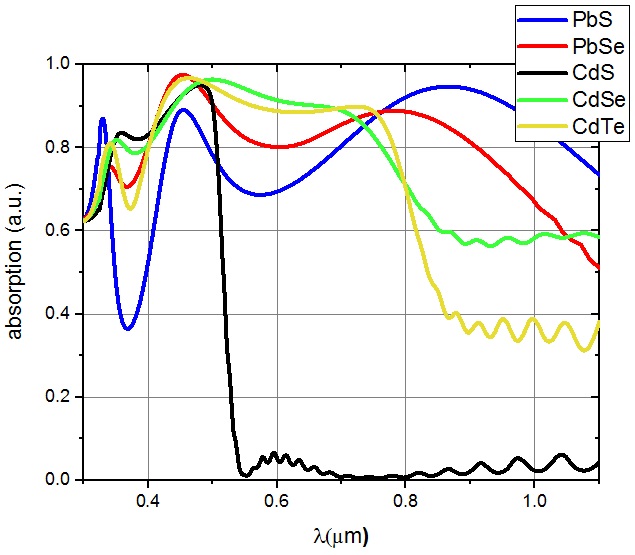}
 \caption{Absorption spectrum of the active area of solar cell.\label{p1}}
\end{figure}

We performed the calculations to obtain the absorption spectrum of cells according to the described equations and the absorption spectrum of QDSSCs with different quantum dots in wavelengths between 300 nm and 110 nm are shown in Fig. \ref{p1}. Among these nanomaterials, \ce{PbS} and \ce{PbSe} can greatly increase the rate of electron-hole production due to the expansion of the absorption band towards the infrared region. Thus, the current density of cells using these materials should be more than other cells. 

\begin{figure}
\centering
\includegraphics[scale=.5]{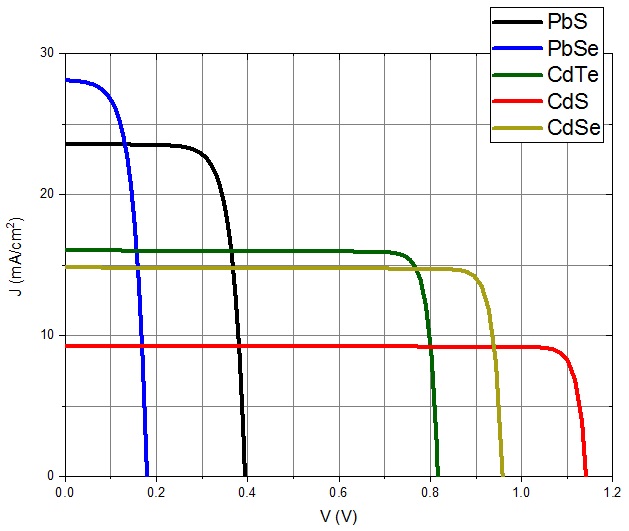}
 \caption{Current density - voltage curves of QDSSCs.\label{p2}}
\end{figure}

As can be seen in the diagram of the current density-voltage of these cells (see Fig. \ref{p2}), the \ce{PbSe} and \ce{PbS} quantum dots have low $V_{OC}$ despite their very good light absorption, due to the low band-gap. The \ce{CdTe}, \ce{CdS} and \ce{CdSe} quantum dots have higher $V_{oc}$ due to the higher band-gap and thus the accumulation of more energetic electrons. 

The cells characterization results are summarized in Table \ref{tabl}, which shows that \ce{CdSe} has the highest efficiency.

\begin{table}
\caption{Results of characterization of current density-voltage curves.}\label{tabl}
\vspace*{0.4 cm}
\centering
\begin{tabular}{lllll}
\hline\noalign{\smallskip}
$sensitizer$&$J_{SC}(mA/cm^2)$&$V_{OC} (V)$&$\eta(\%)$&$F\hspace{-0.5mm}F$\\
\noalign{\smallskip}\hline\noalign{\smallskip}
\ce{PbS}&$23.6$&$0.4$&$7.7$&$0.8$\\

\ce{PbSe}&$28.1$&$0.2$&$3.1$&$0.5$\\

\ce{CdS}&$9.3$&$1.1$&$9.5$&$0.9$\\

\ce{CdSe}&$14.8$&$1.0$&$13$&$0.8$\\

\ce{CdTe}&$16.1$&$0.8$&$12$&$0.9$\\
\noalign{\smallskip}\hline
\end{tabular}
\end{table}

\subsection{Co-sensitized solar cells}
\begin{figure}
\centering
\includegraphics[scale=.5]{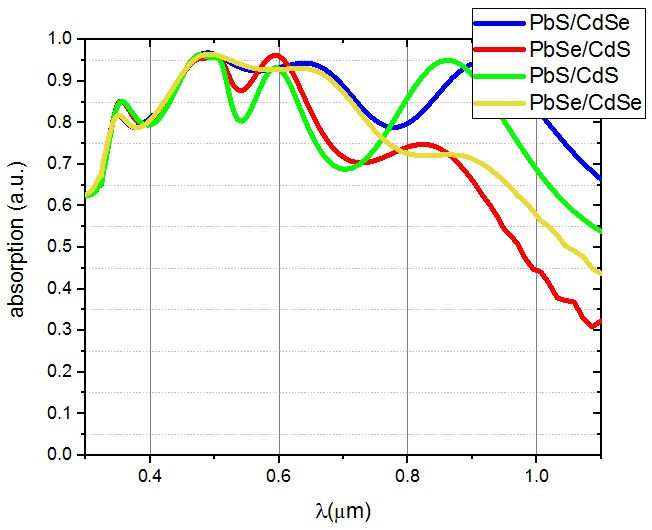}
 \caption{Absorption spectrum of co-sensitized solar cells. \label{p3}}
\end{figure}

By using two types of QDs, the absorption and efficiency of the solar cell can be controlled by two substances. We performed the calculations to obtain the absorption spectrum of co-sensitized solar cells with different pairs of QDs, and the absorption spectrum of them in wavelengths between 300 nm and 110 nm are shown in Fig. \ref{p3}. By comparing Fig. \ref{p1} and Fig. \ref{p3}, it can be concluded that using two different sensitizers, one of them is selected from compounds with small band-gap and the other from compounds with larger band-gap, it increases the absorption of the cell.

\begin{figure}
\centering
\includegraphics[scale=.45]{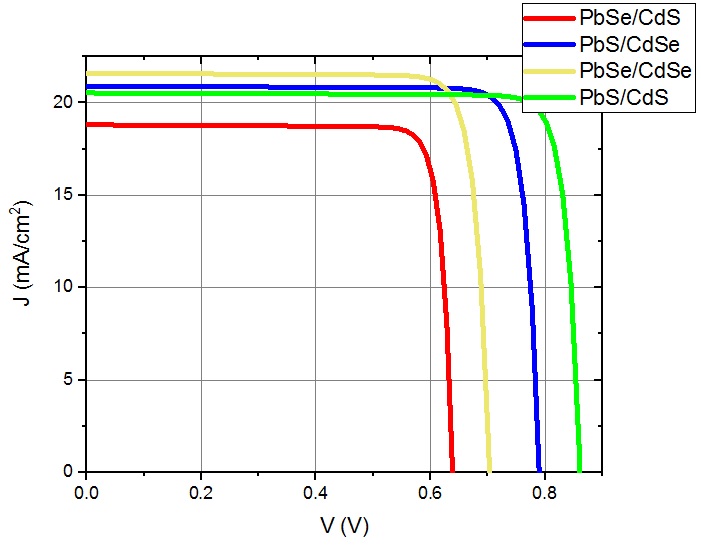}
 \caption{Current density - voltage curves of co-sensitized solar cells.\label{p4}}
\end{figure}

It has been shown experimentally that in the presence of polysulfide electrolyte, photocurrent of sensitized \ce{TiO2} single crystals by \ce{PbS} has a quantum efficiency greater than one electron per photon \cite{li2013efficient}. However, as mentioned, the cell made only with this material has low $V_{oc}$. The characterization results for co-sensitized cells are summarized in Table \ref{tab2}, which shows that the \ce{PbS }/ \ce{CdS} sensitized solar cell has higher $V_{oc}$ than \ce{PbS} sensitized solar cell. Fig. \ref{p4} shows the current density-voltage of cells sensitized by two types of quantum dots.

In co-sensitized solar cells that one type of QDs is selected from the \ce{PbS} and \ce{PbSe} compounds to increase the rate of carrier production by extending the absorption range to the infrared region. The other type is selected from compounds by band-gap more than 1 eV to improve the $V_{OC}$ in addition to increasing the absorption at shorter wavelengths, which can achieve higher efficiency than sensitized solar cells with one type of quantum dot. Our simulation results show the highest efficiency for a sensitized cell using \ce{PbS} and \ce{CdS}. So we chose this cell for the experimental section.

\begin{table}
\caption{Results of characterization of current density-voltage curves of co-sensitized solar cells.}\label{tab2}

\begin{tabular}{llllll}
\hline\noalign{\smallskip}
$sensitizers$&$J_{SC}(mA/cm^2)$&$V_{OC} (V)$&$\eta(\%)$&$F\hspace{-0.5mm}F$\\
\noalign{\smallskip}\hline\noalign{\smallskip}
\ce{PbSe / CdS}&$18.8$&$0.6$&$10.4$&$0.9$\\

\ce{PbS / CdSe}&$20.9$&$0.8$&$14.3$&$0.9$\\

\ce{PbSe / CdSe}&$21.6$&$0.7$&$13.0$&$0.8$\\

\ce{PbS / CdS}&$20.5$&$0.9$&$15.5$&$0.8$\\
\noalign{\smallskip}\hline
\end{tabular}
\end{table}

In the all simulations,  quantum dots are considered dome-shaped Which randomly with different sizes are  arranged on the \ce{TiO2}. The refractive index for QDs can be calculated using Eq. (\ref{qd1}) that N in this equation is defined as $N = n + in_{i}$. $n$ and $n_{i}$ are real and imaginary part of refractive index \cite{sabaeian2012size}.

\begin{equation}
N = \sqrt{1 + \chi_{eff}(\omega)} \approx {1 + \frac{1}{2} \chi_{eff}(\omega)}, \label{qd1} 
\end{equation}

\noindent where $\chi_{eff}(\omega)$, is defined as $\chi_{eff}(\omega) = \chi^{(1)}+ \chi^{(2)}(\omega)\tilde{E} + \chi^{(3)}(\omega)\tilde{E}^{2} $.  $\chi^{(1)}$, $\chi^{(2)}$ and $\chi^{(3)}$ are linear, second order, and third order susceptibilities, respectively. $\tilde{E}$ is a linear x-polarized monochromatic electric field propagate along the z direction (see Eq. (\ref{qd2})) \cite{sabaeian2012size}.

\begin{equation}
\tilde{E}(z,t) = E_0 \hat{i} e^{i(kz-\omega t)}+ C.C . \label{qd2} 
\end{equation}

The relative refractive index change and absorption coefficient of medium can be calculated using Eqs. (\ref{qd3}) and (\ref{qd4}) \cite{vahdani2010intersubband}.

\begin{align}
\frac{\Delta n}{n} = \frac{n-1}{n} = \frac{1}{2} Re(\frac{\chi_{eff}(\omega)}{n})\label{qd3},\\
\alpha = \frac{2 n_i \omega}{c}= \omega \sqrt{\dfrac{\mu}{\epsilon_R} Im[\epsilon_0 \chi_{eff}(\omega)]},\label{qd4}
\end{align}
\noindent where $\mu$ is the vacuum permeability and $\epsilon_R$ is the real part of permittivity. 



\begin{figure}
\centering
\includegraphics[scale=.6]{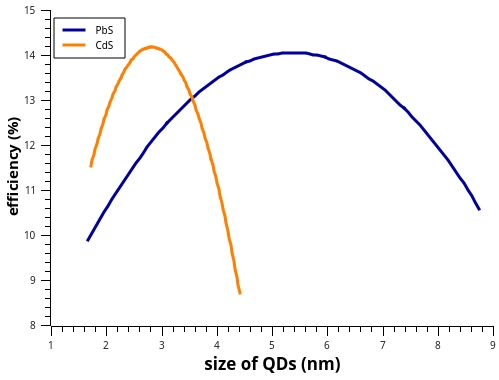}
 \caption{Relation between QDs size and efficiency of solar cell.\label{pq2}}
\end{figure}

After that we simulated QDs of a certain size instead of different sizes that are randomly positioned on \ce{TiO2} and optimized the size of QDs for \ce{PbS}/\ce{CdS} co-sensitized solar cell. Fig. \ref{pq2} shows the results of this optimization and we see that the highest efficiency is obtained with 5.5 nm \ce{PbS} QDs and 2.5 nm \ce{CdS} QDs.

\section{Experimental methods}
\subsection{Anode preparation }
FTO (Fluorine-Tin-Oxide (FTO) Coated Glass Plates; sheet resistance $8 \hspace{0.5 mm}\Omega $/{ $square $}) were used as substrates and pretreated by rinsing in an ultrasonic bath of detergent and acetone for 10 and 20 minutes in turn, then rinsing with a large amount of deionized water and ethanol, and at last dried with nitrogen gas. Transparent titanium dioxide (TiO2) paste was applied to substrates using the doctor-blade method. \ce{TiO2} nanoparticles were in the anatase phase, with an average size of $20-25 nm$. The nanoporous nature of the \ce{TiO2} layer provides a high surface area, which can increase the QDs adsorption and then high photocurrent generation. After the application of transparent \ce{TiO2} paste, the substrates were heated at $120^\circ ~C$ for 0.5 h. Reflector anatase \ce{TiO2} paste after that was applied on the previous layer by the same method, and the deposited photoanodes were heated at $450^\circ ~C$ for 0.5 h. This material provides a high scattering layer as a reflector that increases light harvesting and photocurrent. 



For deposition of \ce{PbS} and \ce{CdS} quantum dots on \ce{TiO2} electrode we used successive ionic layer adsorption and reaction (SILAR) method. The electrode first dipped in a 0.02 M \ce{Pb(NO3)2} aqueous solution for 30 s, rinsed with deionized water and then dipped in 0.02 M \ce{Na2S} aqueous solution for another 30 s followed by rinsing which was termed as one SILAR cycle. The next step is the deposition of \ce{CdS} quantum dots. Similarly, for the \ce{CdS} nanoparticles layer, the electrode first dipped in a 0.2 M \ce{Cd(NO3)2} aqueous solution for 5 min, rinsed and then dipped in 0.2 M \ce{Na2S} aqueous solution for another 5 min. 


The number of immersion cycles can control the size of quantum dots.  The method is designed to increase the particle size of a monolayer during an immersion cycle. The optimal number of cycles can be obtained by repeating the experiment several times to maximize cell efficiency. In this study, layers with different cycles were used to sensitize the \ce{TiO2} electrode once only with \ce{PbS} quantum dots, once only with \ce{CdS} quantum dots and then by combining two types, with \ce{PbS}/\ce{CdS} order as hybrid samples. Four sensitized electrode specimens were thus obtained for use in cell fabrication.

\subsection{Electrolyte preparation}
In dye-sensitized solar cells, platinum is typically used as the counter electrode and iodide/triiodide as the oxidation/reduction pair, but unfortunately, the well-known pair of $I^{-}/I_{3}^{-}$ is not compatible with low band-gap semiconductors and lead to rapid corrosion of these materials \cite{ruhle2010quantum}. For QDSSCs, other pairs such as cobalt compounds, ferrocene/ferrocenium, and polysulfide seem to work well\cite{sapp2002substituted,tachibana2008performance}. Although cobalt compounds are suitable for low energy band-gap semiconductors, this is good for low light intensity cases and has a negative effect on cell function in high light intensity cases.

Polysulfide electrolyte were prepared by mixing suitable quantities of \ce{Na2S}, \ce{S} and \ce{KCl} powders in 3:7 water/methanol solution. Then we put it on the magnetic stirrer for two hours to obtain a uniform solution. The color of the solution changes from yellow to orange after stirring. 


\subsection{Counter electrode}
Different materials and different deposition methods were used to prepare the counter electrodes. For this purpose, the \ce{FTO} is being washed in the same way as in the preparation of the substrates described above. The method for deposition of \ce{Pt} and \ce{Au} was physical vapor deposition (PVD) at the pressure of $10^{-5}$ millibars, and the thickness of the deposited layer was of 100 nm. Another approach was to use \ce{CoS} nanoparticles, which we applied to the pre-prepared \ce{FTOs} using the SILAR method. The counter electrodes were prepared by first dipping a clean FTO in 0.5 M \ce{Co(CH3COO)2} aqueous solution for 30 s, rinsed with deionized water, and then dipped in 0.5 M \ce{Na2S} aqueous solution for another 30 s followed by rinsing. This constituted one cycle, and the process was repeated four times.

The sensitized \ce{TiO2} electrode was then combined with the counter electrode in the presence of polysulfide electrolyte to assemble a typical QDSSC. Photocurrent-voltage (I-V) characteristics of the QDSSCs were measured using a Keithley 2400 electrometer under illumination from xenon lamp at the intensity of $100 ~ mW/cm^2$.

\subsection{Determination of optimal SILAR cycles for deposition of \ce{PbS} and \ce{CdS} quantum dots}
\begin{figure}
\centering
\includegraphics[scale=.5]{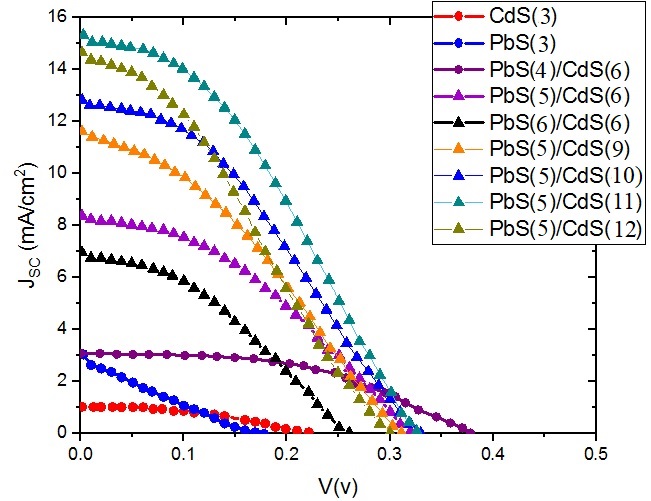}
 \caption{Comparison of the current density - voltage curves of sensitized solar cells with \ce{PbS} and \ce{CdS} quantum dots with different SILAR cycles.\label{p8}}
\end{figure}

The prepared \ce{TiO2} layer electrodes were sensitized using \ce{PbS} and \ce{CdS} quantum dots during different cycles. We then integrated these electrodes with the \ce{Pt} counter electrodes, and after injecting the electrolyte into them, we characterized the fabricated cells. Fig. \ref{p8} shows the voltage-current density of experimental samples that are sensitized by these two types of quantum dots.

Characteristics of these samples, such as efficiency, open-circuit voltage, and short circuit current density and filling factor, were also measured and calculated, as shown in Table \ref{tab5}.

\begin{figure}
\centering
\includegraphics[scale=.65]{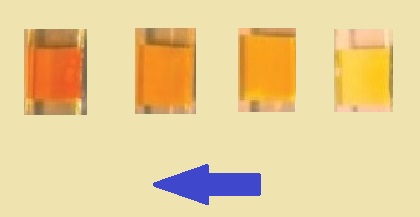}
 \caption{As the number of SILAR cycles increases, the color of the electrode changes from pale yellow to orange.\label{p9}}
\end{figure}

As the number of SILAR cycles in \ce{PbS} deposition changes, the color of the photoelectrodes changes from pale brown to black. According to the results in Table \ref{tab5}, the number of optimal cycles in the deposition of \ce{PbS} is 5, and in the deposition of \ce{CdS} is 11 for which the maximum cell efficiency was obtained. With the increasing number of SILAR cycles for \ce{CdS}, the color of the electrodes changes from pale yellow to orange (see Fig. \ref{p9}).

\begin{table}
\caption{Results of current density-voltage characterization.}\label{tab5}
\begin{tabular}{llllll}
\hline\noalign{\smallskip}
\ce{PbS(X)}&\ce{CdS(Y)}&$J_{SC}(mA/cm^2)$&$V_{OC} (V)$&$\eta(\%)$&$F\hspace{-0.5mm}F$ \\
\noalign{\smallskip}\hline\noalign{\smallskip}
$0$&$3$&$1.13$&$0.23$&$0.12$&$0.45$\\

$3$&$0$&$3.00$&$0.18$&$0.13$&$0.24$\\

$4$&$6$&$3.01$&$0.39$&$0.59$&$0.51$\\

$5$&$6$&$8.44$&$0.33$&$1.06$&$0.38$\\

$6$&$6$&$7.03$&$0.26$&$0.69$&$0.38$\\

$5$&$9$&$11.65$&$0.31$&$1.26$&$0.35$\\

$5$&$10$&$12.81$&$0.33$&$1.56$&$0.37$\\

$5$&$11$&$15.28$&$0.33$&$1.88$&$0.38$\\

$5$&$12$&$14.65$&$0.31$&$1.42$&$0.32$\\
\noalign{\smallskip}\hline
\end{tabular}
\end{table}

The results show that in photoanode sensitized with quantum dots by the SILAR method, as the number of solar cycles increases, the short-circuit current density, open-circuit voltage, and subsequently the cell efficiency increase until the optimized cycles, and then they decrease. When quantum dot deposition is performed with a low number of SILAR cycles, the quantum dots cover the \ce{TiO2} layer partially, and an increase in the number of deposition cycles results in complete \ce{TiO2} layer coverage. If sensitization is performed over an optimal number of cycles, this increase in the cycle reduces cell performance, which can be attributed to poor charge injection and appears to be due to reduction of the quantum effect of larger quantum dots, which reduces the repulsive force for charge injection (electrons and holes) \cite{els7}, increasing the number of recombinant traps in large quantum dots \cite{els6} and prevention of oxidation/reduction ions transfer due to blocking of pores of the structure by quantum dots \cite{jung2012zns}.

\subsection{Investigation of the effect of different counter electrodes}
\begin{figure}
\centering
\includegraphics[scale=.4]{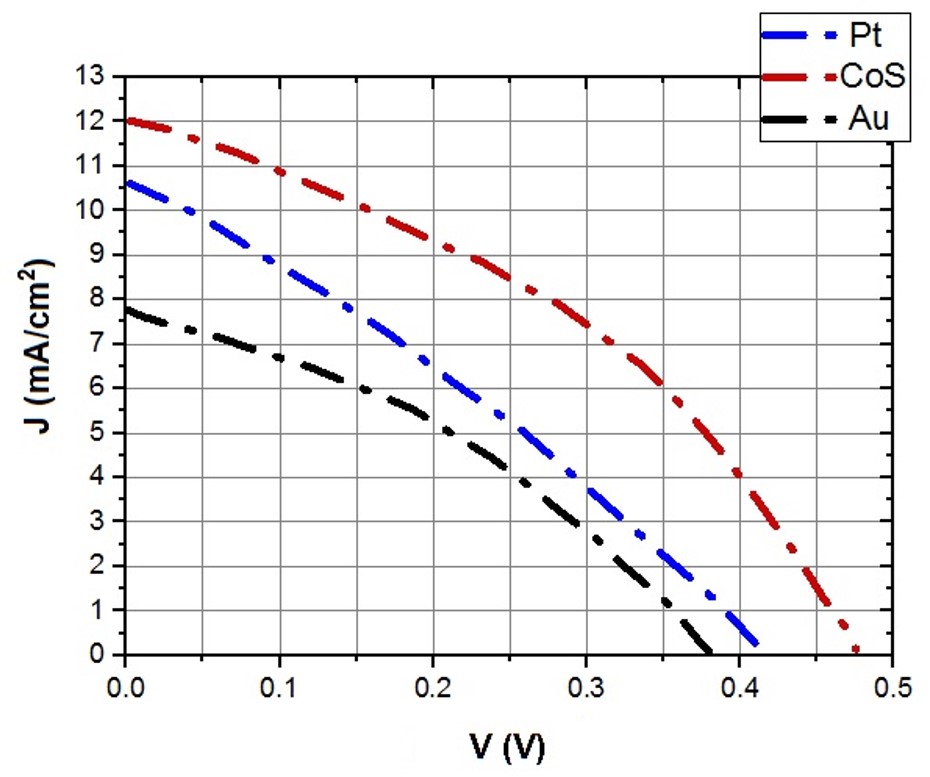}
 \caption{Current density - Voltage curves for solar cells sensitized with \ce{PbS} and \ce{CdS} quantum dots in the presence of \ce{Au}, \ce{Pt} and \ce{CoS} electrodes.\label{p10}}
\end{figure}

Generally, the efficiency of QDSSCs is below 5\%. But simulation and theory studies have shown that the efficiency of these cells can reach up to 44\% \cite{hanna2006solar}. The recombination of carriers at the boundary of the electrolyte quantum dots and the low activity of some of the counter electrodes in the presence of some electrolytes are two major reasons for this low efficiency \cite{fan2009highly}. Gold and platinum are commonly used as counter electrodes and polysulfide electrolytes in QDSSCs. But the conductivity of these electrodes is reduced by the absorption of electrolyte sulfur atoms \cite{miller1976semiconductor}. Therefore, metal sulfides such as \ce{CoS} are suitable substitutes for \ce{Au} and \ce{Pt}, which in addition to having high electrocatalytic activity, can transfer holes much better than these materials and improve cell efficiency \cite{yang2010electrocatalytic}. The Hades Research Group showed that the short-circuit current density is $60 ~ mA/cm^2$ and $20 ~ mA/cm^2$ in the presence of polysulfide electrolytes for \ce{CoS} and \ce{Pt}, respectively \cite{hodes1980electrocatalytic}. We investigated a QDSSC with \ce{PbS} and \ce{CdS} sensitizers for the three different electrodes \ce{Pt}, \ce{Au} and \ce{CoS} which current-voltage-density diagram is shown in Fig. \ref{p10}.

Table \ref{tab20} shows the results of the characterization of these three cell types. Power conversion efficiency using \ce{CoS} increased by about 70\% compared to \ce{Pt} and by about 100\% compared to \ce{Au}. By comparing the current-voltage density diagrams of these three types of cells, it can be seen that the electrodes based on these two metals react strongly with sulfide ions, thereby greatly reduce their catalytic activity and conductivity. However, the \ce{CoS} electrode is more stable and efficient in the presence of polysulfide electrolyte, and its cost is less than the other two types.

\begin{table}
\caption{Results of current density-voltage characterization of cells made with different counter electrodes.}\label{tab20}
\vspace*{0.4 cm}
\centering
\begin{tabular}{lllll}
\hline\noalign{\smallskip}
$counter electrode$&$J_{SC}(mA/cm^2)$&$V_{OC} (V)$ &$\eta(\%)$&$F\hspace{-0.5mm}F$\\
\noalign{\smallskip}\hline\noalign{\smallskip}
\ce{Au}&$7.76$&$0.38$&$1.03$&$0.34$\\

\ce{Pt}&$10.61$&$0.41$&$1.31$&$0.30$\\

\ce{CoS}&$12.01$&$0.47$&$2.23$&$0.39$\\
\noalign{\smallskip}\hline
\end{tabular}
\end{table}

\subsection{absorption spectrum of Photoanode made by optimized number of SILAR cycles}
\begin{figure}
\centering
\includegraphics[scale=.5]{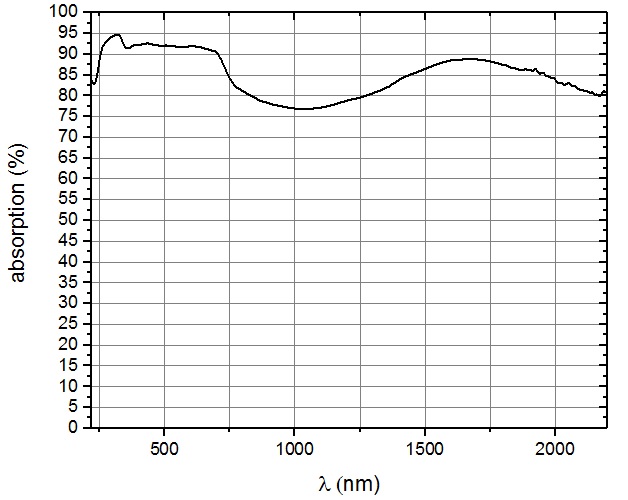}
 \caption{Optical absorption curve of a QDSSC with \ce{PbS} and \ce{CdS} quantum dots made using SILAR method by optimized SILAR cycles.\label{p11}}
\end{figure}

Fig. \ref{p11} shows the absorption spectrum of anode made up of 5 cycles for \ce{PbS} quantum dots and 11 for \ce{CdS} quantum dots. The cell absorbs more than 80\% throughout the sun's spectrum. With high absorption of infrared spectra, this anode can have a high carrier production rate.

Using X-ray diffraction characterization of the optimized cell anode, the size of the \ce{CdS} quantum dots for the optimized cell is approximately 2 nm, and the size of the \ce{PbS} quantum dots was about 5 nm. These magnitudes were obtained by using the Debye-Scherrer relation (Eq. \ref{7eq1}) \cite{yang2005wet}:

\begin{equation}\label{7eq1}
D=\dfrac{0.94\lambda}{\beta cos(\theta)},
\end{equation}

\noindent where $\lambda$ is the wavelength x-ray,  $\beta$ is FWHM and $\theta$ is a diffused Bragg's angle.

\section{Conclusion}
Quantum dot sensitized solar cells with one type of material can achieve more than 12 \% efficiency using \ce{CdSe}. Due to its good band-gap for use in solar cells, it can absorb a wide range of the visible spectrum. Materials with low energy band-gap such as \ce{PbS} and \ce{PbSe} due to their low band-gap have high absorption in the visible and infrared spectrum of light, but cells made with these materials have a smaller $V_{oc}$ because of more recombination of carriers than other materials. Simultaneous sensitizing using two different materials, one with a large band-gap and the other with a small band-gap, can cause the solar cell to absorb more light in the visible, infrared, and ultraviolet spectrum. Although at co-sensitized solar cells, the $V_{OC}$ decreases, overall, the efficiency of this type of cell is higher than a sensitized cell with one type of quantum dot. The co-sensitized cell with the \ce{PbS} and \ce{CdS} quantum dots has a higher efficiency (Almost 15 \%) than the other types of co-sensitized cells due to its $J_{SC}$ higher than $20 ~ mA/cm^{2}$ and $V_{OC}$ higher than $0.8 ~ V$. The experimental method of fabricating solar cells can greatly influence the characteristics of this cell by changing the number of SILAR cycles. The X-ray diffraction results show that by changing the concentration of precursors, the immersion time, and the number of cycles, the size of the quantum dots can be well controlled. A cathode made of \ce{CoS}, which costs less than platinum and gold cathodes, can increase relative cell efficiency of \ce{PbS}/\ce{CdS} co-sensitized solar cell up to 80 \%. Although the anode made with \ce{PbS} and \ce{CdS} quantum dots has high absorption, the efficiency of this cell is low, which can be a result of the strongly recombined carriers in this cell.

\section{Compliance with ethical standards}
\textbf{Conflict of interest} The authors declare that they have no conflict of interest.


%
%

\bibliographystyle{spphys}       
\bibliography{ref}  

%
%


\end{document}